\documentclass[
reprint,
 amsmath,amssymb,
 aps, physrev,
pra,
]{revtex4-2}

\usepackage{graphicx}
\usepackage{dcolumn}
\usepackage{bm}

\usepackage{float}
\begin{document}

\preprint{}

\title{\textbf{Interface-tuned Enhanced and Low Temperature Quenching of Orbital Hall Currents Induce Torque and magnetoresistance in Light Metal/Nickel Bilayers.} 
}%

\author{Dhananjaya Mahapatra}
\email{Contact author: dm20rs019@iiserkol.ac.in}
 \author{Harekrishna Bhunia}
  \author{Manu S Pattelath}
  \author{Partha Mitra}
  \email{Contact author: pmitra@iiserkol.ac.in}
 
\affiliation{%
 Department of Physical Sciences,
 Indian Institute of Science Education and Research Kolkata, Mohanpur, West Bengal,741246, INDIA.
}%




\begin{abstract}
We investigate orbital current induced effects arising from the orbital Hall effect in light-metal/ferromagnet bilayers. Thin films of Ti in ohmic contact with Ni were studied using second-harmonic longitudinal and transverse voltage measurements under an applied a.c. current. From these signals, we extract the orbital Hall torque (OHT) efficiency and the unidirectional orbital magnetoresistance (UOMR). Insertion of a Cu interlayer between the Ni/Ti interface leads to an enhancement of both OHT efficiency and UOMR compared to both Ni/Ti and Ni/Cu bilayers. Furthermore, systematic variation of Ti thickness reveals that both OHT efficiency and UOMR increase with increasing Ti thickness, indicating that the observed phenomena predominantly originate from the bulk orbital Hall effect rather than purely from interfacial mechanisms and Lowering the temperature leads to a clear reduction in both the orbital Hall torque (OHT) efficiency and the unidirectional orbital magnetoresistance (UOMR). The nearly linear and correlated temperature dependence of both parameters suggests a common underlying mechanism, namely, the orbital Hall effect in the light-metal layer, which governs both the generation of orbital current and its subsequent influence on the ferromagnet through orbital torque and orbital magnetoresistance.

\end{abstract}

\maketitle


\section{\label{sec:level1}INTRODUCTION:}

A current flowing through heavy metals (HM) with appreciable spin-orbit coupling (SOC) strength generates transverse pure spin current due to the phenomenon of spin Hall effect (SHE) \cite{PhysRevLett.83.1834, PhysRevLett.85.393, PhysRevLett.92.126603,6516040}, which is confirmed through numerous experiments\cite{15539563, JOUR04937, PhysRevLett.119.087203}. A standard experimental technique that manifests SHE induced spin currents is the measurement of second harmonic voltages in response to an a.c. current flowing through the plane of bilayer heterostuctures HM/FM, where FM is a layer of ferromagnetic conductor\cite{PhysRevB.90.224427}. The polarization axis of the spin current generated in the HM due to SHE is linked only to the direction of current flow which remains constant, while the magnetisation of the FM is rotated continuously with the help of an external magnetic field.
The second harmonic Hall voltage is shown to be linked to magnetisation dynamics in the FM induced by the transfer of the component of spin angular momentum from the injected spin currents arising in the adjacent HM layer, perpendicular to the magnetization of the FM. The effect of the component of the spin current along the magnetisation also reveals itself in the longitudinal voltage, which is referred to as the unidirectional spin Hall magnetoresistance (USMR)\cite{Avci2015, PhysRevLett.121.087207, 10.1063/1.4935497, 10.1063/1.4983784} and can be grossly understood as a spin valve type effect in the second harmonic response,  with resistance varying from minimum to maximum when the magnetisation is rotated from being parallel to antiparallel with respect to the spin current polarization. 
\\Recently theoretical works have pointed out the fact that the carriers in a conductor also possess orbital angular momentum (OAM) in addition to spin angular momentum and predict a new phenomenon analogous to SHE, where carriers with opposite OAM  are segregated in a direction transverse to current flow and is termed as orbital Hall effect (OHE)\cite{PhysRevLett.121.086602, PhysRevB.98.214405, PhysRevResearch.2.013177}. It is further emphasized that the OHE will be relevant even in conductors with insignificant SOC, classified as Light Metals (LM). Such predictions have prompted a series of experiments on LM/FM bilayer structures on the exact same lines as that of HM/FM bilayers mentioned previously\cite{Lee2021, Lee2021nature, PhysRevResearch.4.033037, PhysRevB.107.134423, Choi2023, Hayashi2023}. However, unlike the spin, the OAM of a carrier cannot directly interact with the magnetisation of the FM.  It is now established that Nickel as the FM layer exhibits special characteristics that can effectively convert the OAM current injected from the LM layer into a spin current\cite{PhysRevResearch.2.033401, Lee2021, Lee2021nature, PhysRevB.107.134423}. Although the exact underlying mechanism is yet to be established, it is speculated in some reports that this unique ability of Ni can be attributed to its large SOC compared to other popular FMs like Py, Co, CoFe etc. that form the basis of most reported works on HM/FM bilayers. Once the OAM current is converted to a spin current, the usual picture of a torque acting on the magnetisation of Ni is applicable and is termed as Orbital Torque due to its origin. Similarly, the resistance of LM/Ni bilayers exhibits a change in resistance as the magnetisation orientation is flipped laterally with respect to current flow and the phenomenon is termed as unidirectional orbital magnetoresistance (UOMR)\cite{PhysRevResearch.4.L032041}.
\\
Here, we present a systematic investigation of orbital angular momentum (OAM) current in a series of Hall bar–shaped ferromagnet/light metal (FM/LM) bilayers. We have chosen Ti and Cu as the light metals since both possess negligible spin Hall conductivity but exhibit significant orbital Hall conductivity, as confirmed by recent experimental reports. Nickel (Ni) is used as the ferromagnetic layer, which we have previously shown to effectively convert orbital angular momentum into spin angular momentum. In the present study, a Cu interlayer is inserted between Ti and Ni to examine its influence on orbital transport. Furthermore, we have simultaneously measured the second-harmonic longitudinal and transverse resistances, which are expected to originate from the mutually perpendicular components of the injected orbital current with respect to the magnetisation of the ferromagnet. By systematically varying the thicknesses of Ti and the interfacial Cu layer, and following standard analysis methods used for spin-current studies in FM/heavy metal (HM) bilayers, we extract the torque efficiency and unidirectional magnetoresistance (UMR) per unit electric field and establish a clear correlation between them.

\section{\label{sec:level1}Sample Fabrication and Experimental Details:}
All FM/LM devices are patterned in the shape of standard Hall bars with current channel dimensions of $(2\times 30) \mu m^2$ and voltage pickup line widths of $0.5 \mu$ using ebeam lithography ( ELPHY Quantum, Raith ) on Si/SiO2(300nm) substrates with all FM and LM layers deposited for a thickness of 10nm. The FM layer of either Ni or NiFe was first deposited by the thermal evaporation technique at the rate of (0.8-0.9){\AA} /s. Thereafter, the LM layer of Ti  was deposited on top of the FM by DC sputtering at rates of 0.5-0.6\AA /s. Ar ion milling was performed before all deposition to clean the interface and promote good Ohmic contact between FM and LM. Finally, contact pads were patterned on the Hall bar electrodes using UV photolithography and Cr/Au was deposited using ebeam evaporation. X-ray diffraction (XRD) with a Cu-$K_\alpha$(1.54\AA) source confirms fcc structure for Ti. The bare resistivities of each metallic layers were determined using  Van der Pauw technique  and found to be $\rho_{Ti}=80,\rho_{Ni}=130,\rho_{NiFe}=170 \mu\Omega.cm$. 
For harmonic measurements a constant a.c. current $I(t)=I_o sin(2\omega t)$ is sourced through the bilayer  Hall bar with $\omega=2\pi X 13$rad/s using a Keithley 6221A AC-DC current source and the amplitudes of first and second harmonic voltage responses are recorded across both longitudinal ($V_{xx}^\omega, V_{xy}^{2\omega}$) and transverse ($V_{xy}^\omega,V_{xy}^{2\omega}$) voltage leads of the Hall bars using a SR830 Lockin amplifier. The corresponding resistances are defined as the ratio of the voltage amplitude to current amplitude, in particular $R_{xx}^{2\omega}=V_{xx}^{2\omega}/I_o$ and $R_{xy}^{2\omega}=V_{xy}^{2\omega}/I_o$. As the FM and NM layers are in ohmic contact, the applied current is distributed between the two layers depending on the resistivity of the layers and the current density in each layers can be estimated as, 
$J_{LM}$=$\frac{I_{o}}{(wt)[1+(\frac{\rho_{LM}}{\rho_{FM}})]}$ and $J_{FM}$=$\frac{I_{o}}{(wt)[1+(\frac{\rho_{FM}}{\rho_{LM}})]}$. 
The value of $I_o$ is suitably chosen so that the estimated $J_{LM}$ in devices with same LM layer but different FM layers (Ni or NiFe) remains same, ensuring same orbital Hall current. Typical current densities for our experiments are in the range of $10^{11} A/m^2$. Our goal is to investigate the harmonic responses of the FM/LM bilayers as the FM magnetisation is rotated with respect to the polarization of the OAM current arising from OHE in the LM layer.
The devices are mounted on a homemade setup with 2D vector electromagnet ( DEXING, China) with the Hall bar lying in the plane of the magnetic field vector and the current channel along the $x$-component of magnetic field. The measurement of Planar Hall effect confirms that the magnetisation of the FM is indeed rotating with the magnetic field. A typical variation of the second harmonic resistances for a Ni/Ti device from $B=0.4$T   reveals that the variation of $R_{xy}^{2\omega}$ is opposite to that of $R_{xx}^{2\omega}$. As $R_{xy}^{2\omega}$ bears the effect of orbital torque transferred to the perpendicular component of magnetisation from the OAM current, it reaches maximum value at $\Phi=90^o, 270^o$. $R_{xx}^{2\omega}$ on the other hand arises due to GMR like effect where the resistance depends on the relative orientation of OMR polarisation and majority spins of the FM and and reaches extrema for $\Phi=0^o and 180^o$. In the subsequent section we present through analysis of the second harmonic data. \\\\

\section{\label{sec:level1}result and discussion:}
 \begin{figure*}    
\includegraphics[width=1\textwidth]{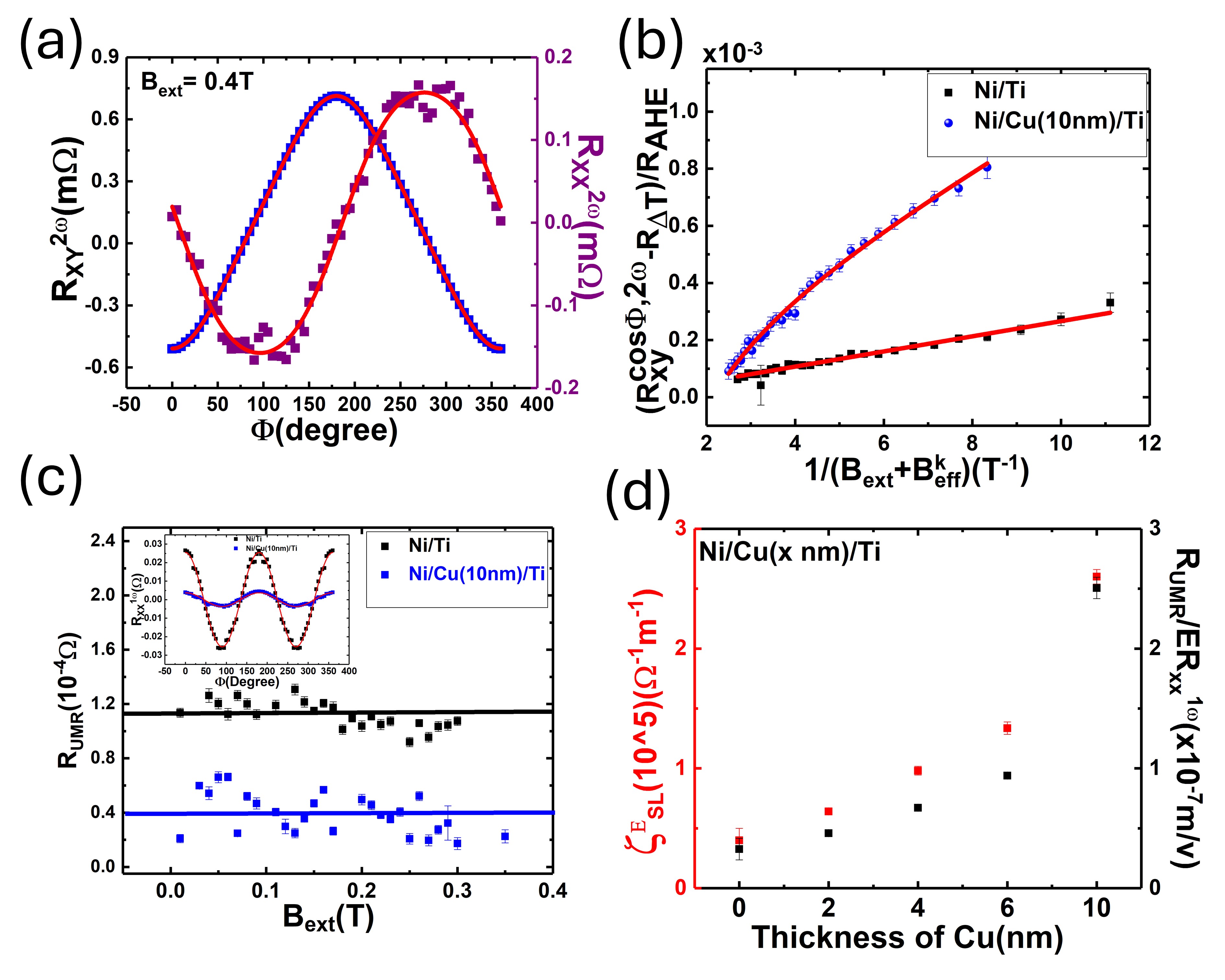}
 \caption{(a).Simultaneous measurements of transverse Hall and longitudinal second-harmonic signals were obtained using a lock-in amplifier of FM/LM bilayer device. (b). Field dependence of the second-harmonic transverse resistance ($\frac{R_{xy}^{cos\phi, 2\omega}-R_{\nabla T}}{R_{AHE}}$) plotted as a function of $\frac{1}{B_{ext}+B^k_{eff}}$ for Ni/Ti and Ni/Cu(1 nm)/Ti bilayers. The slope represents the orbital Hall torque efficiency, which shows a significant enhancement upon Cu insertion. (c). Magnetic-field dependence of the unidirectional magnetoresistance $R_{UMR}$ for Ni/Ti and Ni/Cu/Ti samples; the inset shows the angular dependence of the first-harmonic voltage. (d). Comparison of the orbital torque efficiency($\zeta_{SL}$) and normalized UMR amplitude ($R_{UMR}/ER_{xx}^{1\omega}$) for the two structures, highlighting a correlated $\approx 5\times$ enhancement after Cu insertion, indicating their common origin associated with the orbital Hall effect.}
 \label{schematic}
\end{figure*}
The first harmonic transverse voltage response of the FM/HM bilayers is a linear effect that depends only on  the equilibrium state of magnetization.  The in-plane and out-of-plane components of the magnetization results in Planar Hall Effect(PHE) and Anomalous Hall Effect(AHE) respectively, which contributes to the $R_{xy}^\omega$ depending on the relative angle between magnetization, controlled by the external magnetic field ($B_{ext}$), with respect to current. The second harmonic voltage response arises from precessional motion of the magnetization around the equilibrium position due to toque induced by the spin current\cite{PhysRevB.90.224427,PhysRevB.100.214438} injected from the adjacent HM either due to bulk Spin Hall Effect or Rashba Edelstein type inetrface effects. As the second harmonic voltage is non-linear in current, the corresponding resistance $R_{xy}^{2\omega}$ is  dependent on current it is customary to calculate effective fields per unit current that exerts the spin orbit torques per unit magnetization, also known as torque efficiencies.

\begin{figure*}   
\includegraphics[width=1\textwidth]{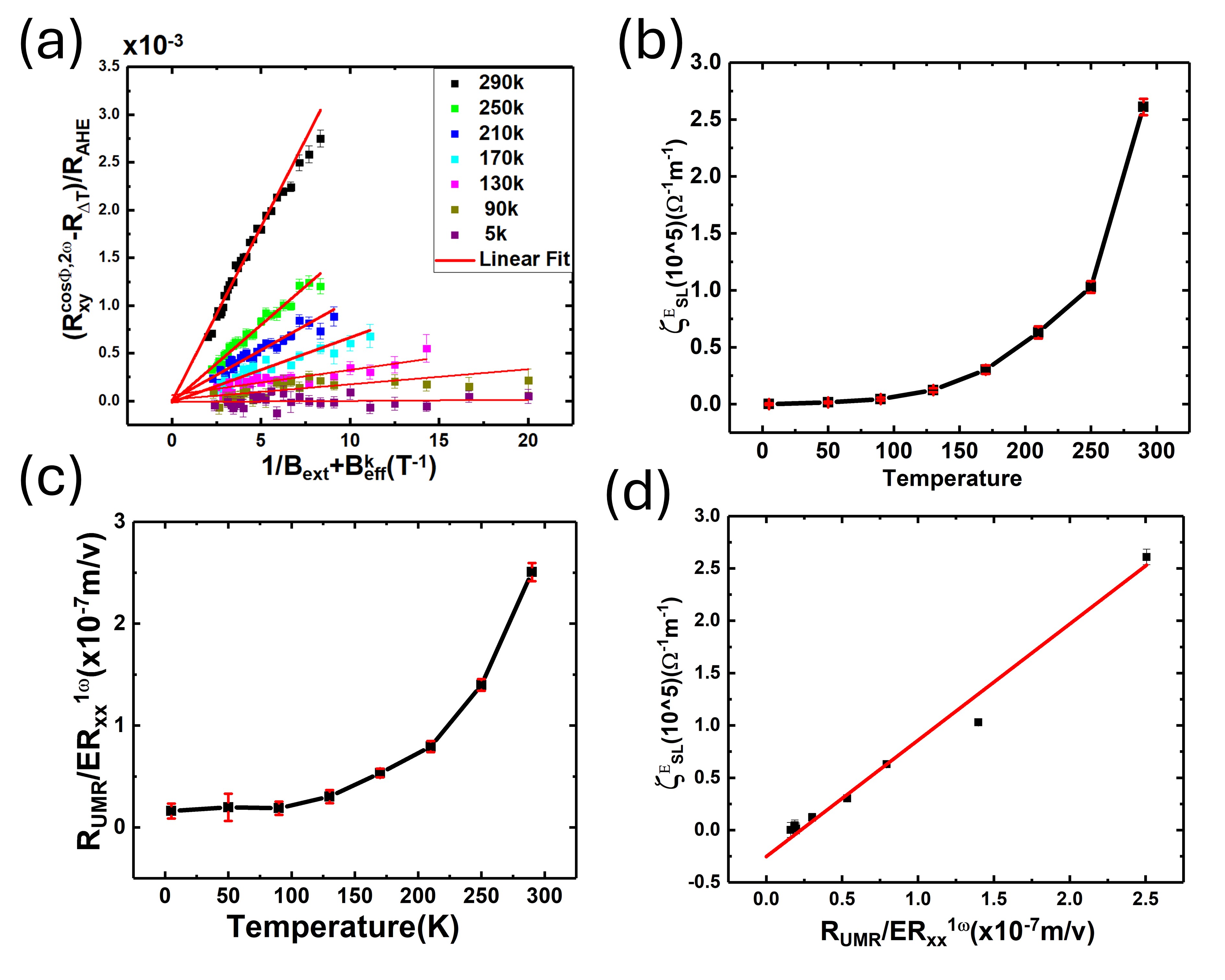}
 \caption{(a). Field dependence of the second-harmonic transverse resistance ($\frac{R_{xy}^{cos\phi, 2\omega}-R_{\nabla T}}{R_{AHE}}$) plotted as a function of $\frac{1}{B_{ext}+B^k_{eff}}$ for the Ni/Cu/Ti bilayer. The slope of the linear fit provides the orbital Hall torque (OHT) efficiency($\zeta_{SL}$). (b).Temperature dependence of $\zeta_{SL}$, showing a monotonic decrease with decreasing temperature. (c). Temperature variation of normalized unidirectional orbital magnetoresistance ($R_{UMR}/ER_{xx}^{1\omega}$), exhibiting a similar trend. (d). Linear correlation between $\zeta_{SL}$ and $R_{UMR}/ER_{xx}^{1\omega}$, indicating that both effects originate from the same underlying mechanism—the orbital Hall effect in the light-metal layer.}
 \label{schematic}
\end{figure*}

In FM/LM bilayers, the LM layer doesnot generate any spin current under an applied current bias due to lack of appreciable Spin-Orbit coupling,  Instead a transverse OAM current $J_{OH}$ is generated and injected into the adjacent FM layer\cite{PhysRevLett.121.086602}. However, unlike the spin currents ,  the orbital current does not directly interact with the  magnetization of the FM and hence will not reveal itself in the form of a torque under ordinary circumstances. Recent findings suggest that certain ferromagnetic materials e.g Ni \cite{Lee2021, Lee2021nature, Choi2023, PhysRevResearch.2.033401} and Co \cite{PhysRevResearch.4.L032041},  can facilitate the conversion of orbital current into spin current with certain efficiency, possibly due its inherent spin-orbit coupling (SOC).  This spin current will induce magnetization dynamics similar to the situation in FM/HM and the resulting torque is termed as Orbital torque and can be analyzed on the same footing as SOT\cite{PhysRevB.100.214438} and the angular dependence of $R_{xy}^{2\omega}$ for FM layers with  in-plane magnetization, is expressed by the phenomenological equation, by\cite{PhysRevB.99.195103, JOUR, AHN202312}
\begin{equation}
\begin{split}
    R_{xy}^{2\omega}(\Phi)=[R_{AHE}(\frac{B_{SL}}{B_{ext}+B_{eff}^k})+\alpha B_{ext} + R_{\nabla T}]cos(\Phi)\\
    +2R_{PHE}(\frac{B_{FL}+B_{Oe}}{B_{ext}})cos(2\Phi)cos(\Phi)
    \end{split}
    \label{eq1}
\end{equation}
where $B_{SL}$ and $B_{FL}$ are the  Slonczewski like and Field-like effective fields that exert in-plane and out-of-plane torque on the magnetisation, respectively.  $R_{AHE}$ is the anomalous Hall resistance, $B_{Oe}$ is the Oersted field , $B_{eff}^{k}$ is the effective anisotropy field of the FM layer. $\alpha$ is the ordinary Nernst coefficient and $R_{{\nabla}T}$ is the anomalous Nernst coefficient, which are spurious heating effect terms due to lthe arge current density required in the experiments.\\\\

 \begin{figure*}  
\includegraphics[width=1\textwidth]{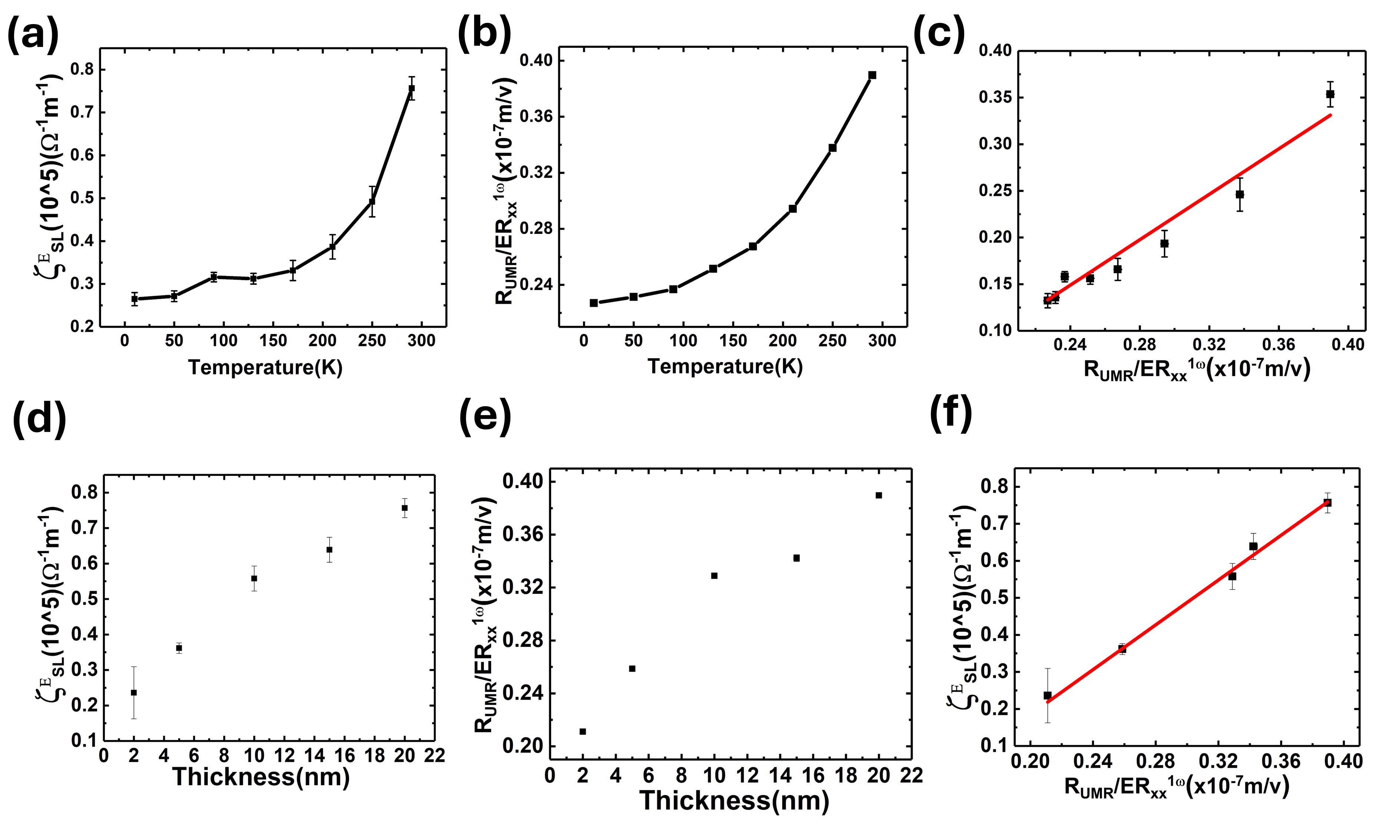}
 \caption{(a).Temperature dependence of the orbital Hall torque efficiency $\zeta_{SL}$ for the Ni/Ti bilayer, showing a gradual decrease with decreasing temperature. (b). Temperature variation of the normalized unidirectional orbital magnetoresistance $R_{UMR}/ER_{xx}^{1\omega}$, exhibiting a similar trend. (c). Linear correlation between  $\zeta_{SL}$ and $R_{UMR}/ER_{xx}^{1\omega}$, indicating that both effects share a common origin arising from the orbital Hall effect in the Ti layer.(d)..Thickness dependence of the orbital Hall torque efficiency $\zeta_{SL}$ for the Ni/Ti bilayer, showing a gradual increase with increasing thickness. (e). Thickness variation of the normalised unidirectional orbital magnetoresistance $R_{UMR}/ER_{xx}^{1\omega}$, exhibiting a similar trend. (f). Linear correlation between  $\zeta_{SL}$ and $R_{UMR}/ER_{xx}^{1\omega}$, indicating that both effects share a common origin arising from the orbital Hall effect in the Ti layer }
 \label{schematic}
\end{figure*}

 In our previous paper\cite{10.1063/5.0263240}, we demonstrated that the Ni/Ti bilayer exhibits a higher orbital torque efficiency as well as a larger orbital unidirectional magnetoresistance (UMR). In this system, the orbital current is primarily generated in Ti and subsequently converted into a spin current within Ni, which exerts a torque — referred to as the orbital torque. For comparison, we also investigated a control sample of NiFe/Ti, where the absence of orbital torque led to a negligible orbital-to-spin conversion efficiency.In the present work, we modify the Ni/Ti interface by introducing a Cu interlayer. Since Cu has a fully filled d band, it serves as an ideal spacer to separate the Ni/Ti interface and allows us to examine the interfacial contribution to orbital current transmission and conversion. We have measured the angular dependence of both the transverse and longitudinal second-harmonic resistance under various external magnetic fields ranging from 0.4 T to 0.02 T. The data were fitted using Eq. \ref{eq1}, and the coefficient corresponding to the $cos\phi$ term was extracted for each magnetic field, as shown in the Supplementary Information. In Fig. 2(a), we plot the extracted $cos\phi$ coefficient (after dividing by $R_{AHE}$ and subtracting the heating-related contribution as a function of $\frac{R_{xy}^{cos\phi, 2\omega}-R_{\nabla T}}{R_{AHE}}$, and the resulting data are fitted using eq of straight line and the slope gives the $B_{SL}$ value.  To compare the orbital torque on different  devices which may be bearing varied current densities, it is customary to calculated the torque efficiency per unit  electric field as\cite{PhysRevB.92.064426}

\begin{figure*}   
\includegraphics[width=1\textwidth]{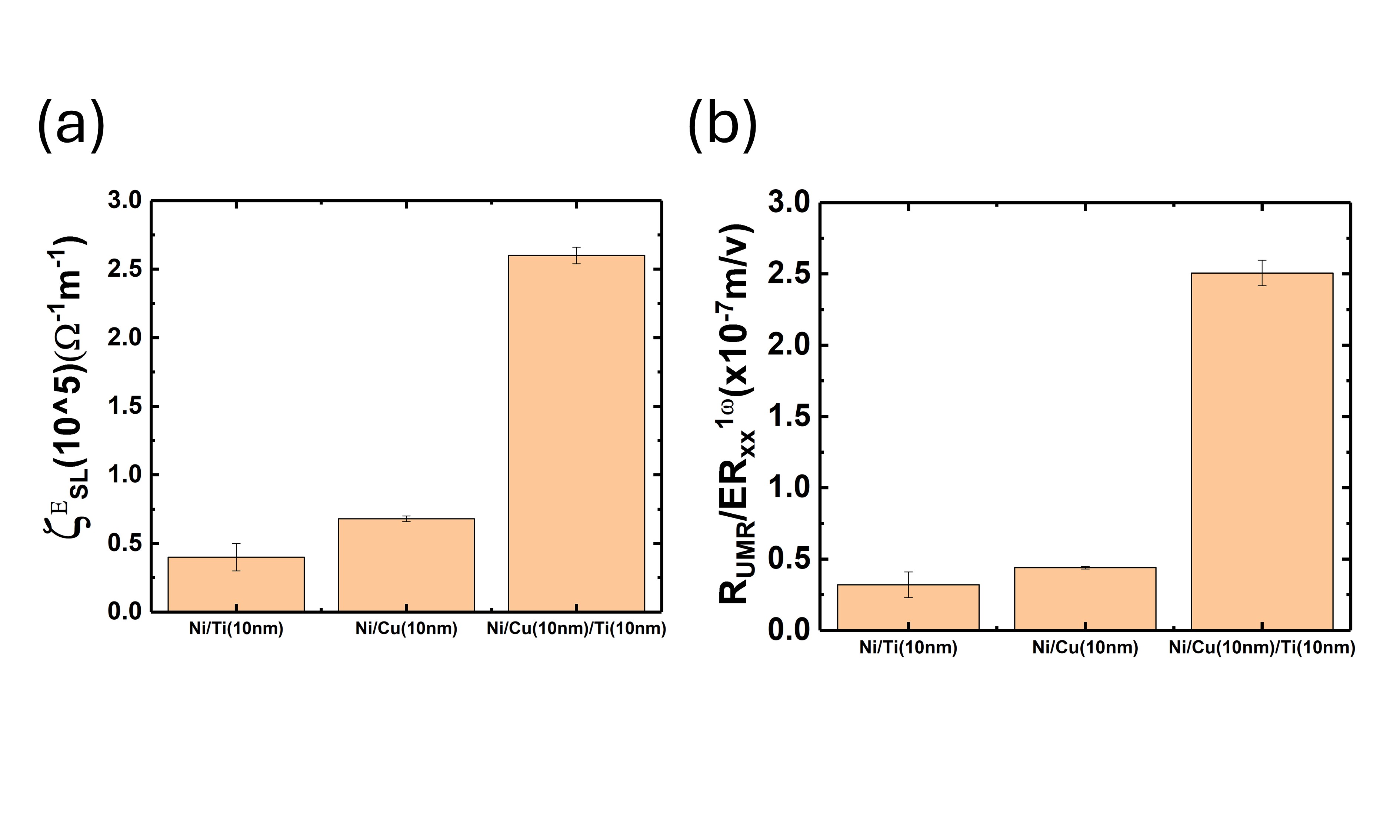}
 \caption{Comparison of (a) the damping-like torque efficiency per unit electric field ($\zeta_{SL}^E$) and (b) the unidirectional magnetoresistance per unit electric field ($R_{UMR}/ER_{xx}^{1\omega}$) for Ni/Ti(10 nm), Ni/Cu(10 nm), and Ni/Cu(10 nm)/Ti(10 nm) bilayers. Both quantities show a significant enhancement in the Ni/Cu/Ti trilayer, indicating efficient orbital-to-spin conversion and a strong correlation between orbital Hall torque and unidirectional orbital magnetoresistance.  }
 \label{schematic}
\end{figure*}

\begin{equation}
    \zeta_{SL}^E=\frac{2e}{\hbar} M_s t_{\small{FM}} \frac{B_{SL}}{E}
\end{equation}
we find $ \zeta_{SL}$ for Ni/Ti and Ni/Ti/Cu(10nm)/Ti  respectvily.
In the context of spin transport in FM/HM bilayer devices, a characteristic $\hat{m}\times\hat{j}$ dependence in the second harmonic response  of longitudinal resistance $R_{xx}^{2\omega}$ is termed as Unidirectional Spin Hall Magnetoresistance (USMR) \cite{Avci2015, PhysRevLett.121.087207, 10.1063/1.4935497, 10.1063/1.4983784}. The analogous phenomenon in a FM/LM system was recently reported \cite{PhysRevResearch.4.L032041} in Cu/Co bilayers and termed as Unidirectional Orbital Magnetoresistance (UOMR).  The underlying mechanism of USMR (or UOMR) is the presence of accumulated spin (or OAM) at the interface of FM with the normal metal, HM(or LM), due to which the overall resistance of the bilayer varies depending on the relative orientation of the FM magnetization with  respect to current direction, similar to CIP GMR phenomenon in FM/NM/FM trilayer devices.  The  resistance of the bilayer is maximum when the majority spins of the ferromagnet is aligned parallel to the accumulated moment, spin (or OAM (Fig. \ref{1}a)) and minimum in the antiparallel configuration (Fig. \ref{1}b).  However, there are additional contribution to the $R_{xx}^{2\omega}$, and the overall angular dependence is found to obey the  expression\cite{PhysRevB.106.094422, PhysRevLett.127.207206}

\begin{equation}
    R_{xx}^{2\omega}(\Phi)=R^* sin(\Phi)-2\Delta R_{xx}^{1\omega} \frac{B_{FL}+B_{Oe}}{B_{ext}} cos^2(\Phi) sin(\Phi)
    \label{equation3}
\end{equation}
The coefficient of $sin(\Phi)(\sim m_y)$ includes the effect of UMR ( either USMR or UOMR) and thermal gradient, $R^*=gR_{\nabla T} + R_{UMR}$. Experimentally, the thermal gradient contribution $R_{\nabla T}$ is independently extracted from the analysis of $R^{2\omega}_{xy}$  and scaled by the  geometric factor  $g = l/w$ , $l$ and $w$ being the length and width of the hall bars respectively. In our devices, $g\approx4$ which is also consistent with the ratio  $\approx \frac{\Delta R^{1\omega}_{xx}}{\Delta R^{1\omega}_{xy}}$. The second term of Eq. \ref{equation3} arises from the effect out of plane torque acting on the magnetization due to combined Oersted and field like effective field and the prefactor $\Delta R_{xx}^{1\omega}$ is the maximum change in the first harmonic resistance as the FM magnetization is flipped from along the current to perpendicular to current configuration.
 In Fig. 2(b), In Fig. 2(b), the $R_{UMR}$ values obtained after subtracting the heating contribution remain nearly constant with varying magnetic fields. Finally, we evaluated both the orbital torque efficiency per electric field and the orbital UMR normalised by the electric field and $R_{xx}^{1\omega}$ signal. A clear fivefold enhancement is observed in both the efficiency and the UMR upon introducing a Cu interlayer at the interface, which likely originates from the modified interfacial properties at the Ni/Cu and Cu/Ti interfaces.\\\\

Next, we measured the temperature dependence of both the Ni/Cu/Ti (Fig. 3) and Ni/Ti (Fig. 4) samples, focusing on the orbital torque efficiency and orbital UMR. In both cases, the efficiency and UMR decrease with decreasing temperature, indicating a similar trend. Finally, we plotted the efficiency and UMR, each normalised by the electric field and the $R_{xx}^{1\omega}$ signal, and observed a linear relationship between them. This linear behaviour suggests a common origin for both the orbital torque efficiency and the orbital UMR in these bilayer structures.
We also investigated the effect of varying the thickness of the light metal (LM) layer. An increase in both the orbital torque efficiency and the orbital UMR was observed with increasing LM thickness, indicating the bulk origin of the orbital current in the LM. Furthermore, the plot of efficiency versus UMR again reveals a linear correlation, reinforcing that both effects share a common origin — the orbital current generated in the light metal layer.

\section{\label{sec:level1}conclusions:}

In summary, we have systematically investigated the influence of interfacial modification, temperature, and thickness on the orbital torque efficiency and orbital unidirectional magnetoresistance (UMR) in Ni/Ti and Ni/Cu/Ti bilayers. Introducing a Cu interlayer at the Ni/Ti interface leads to nearly a fivefold enhancement in both orbital torque efficiency and orbital UMR, which is attributed to the modified interfacial electronic structure at the Ni/Cu and Cu/Ti interfaces. Temperature-dependent measurements reveal that both the efficiency and UMR decrease with decreasing temperature, exhibiting a linear correlation that suggests a common microscopic origin. Similarly, the increase in both quantities with increasing Ti thickness confirms the bulk origin of the orbital current in the light metal. Overall, these results provide compelling evidence that the orbital current generated in the light metal governs both the orbital torque and the orbital UMR, highlighting the crucial role of interface engineering in optimising orbital transport phenomena.

\begin{acknowledgments}
The authors thank IISER Kolkata, an autonomous research and teaching institution funded
by the MoE, Government of India, for providing the financial support and infrastructure. The authors also thank CSIR and UGC for providing fellowship. The authors thank Prof. Bheema Lingam Chittari for valuable discussions.

\end{acknowledgments}
\section*{DATA  AVAILABILITY}
The data that support the findings of this article are not
publicly available. The data are available from the authors
upon reasonable request.
\appendix
\section{Schematic}
\begin{figure}[h!]    
\includegraphics[width=0.5\textwidth]{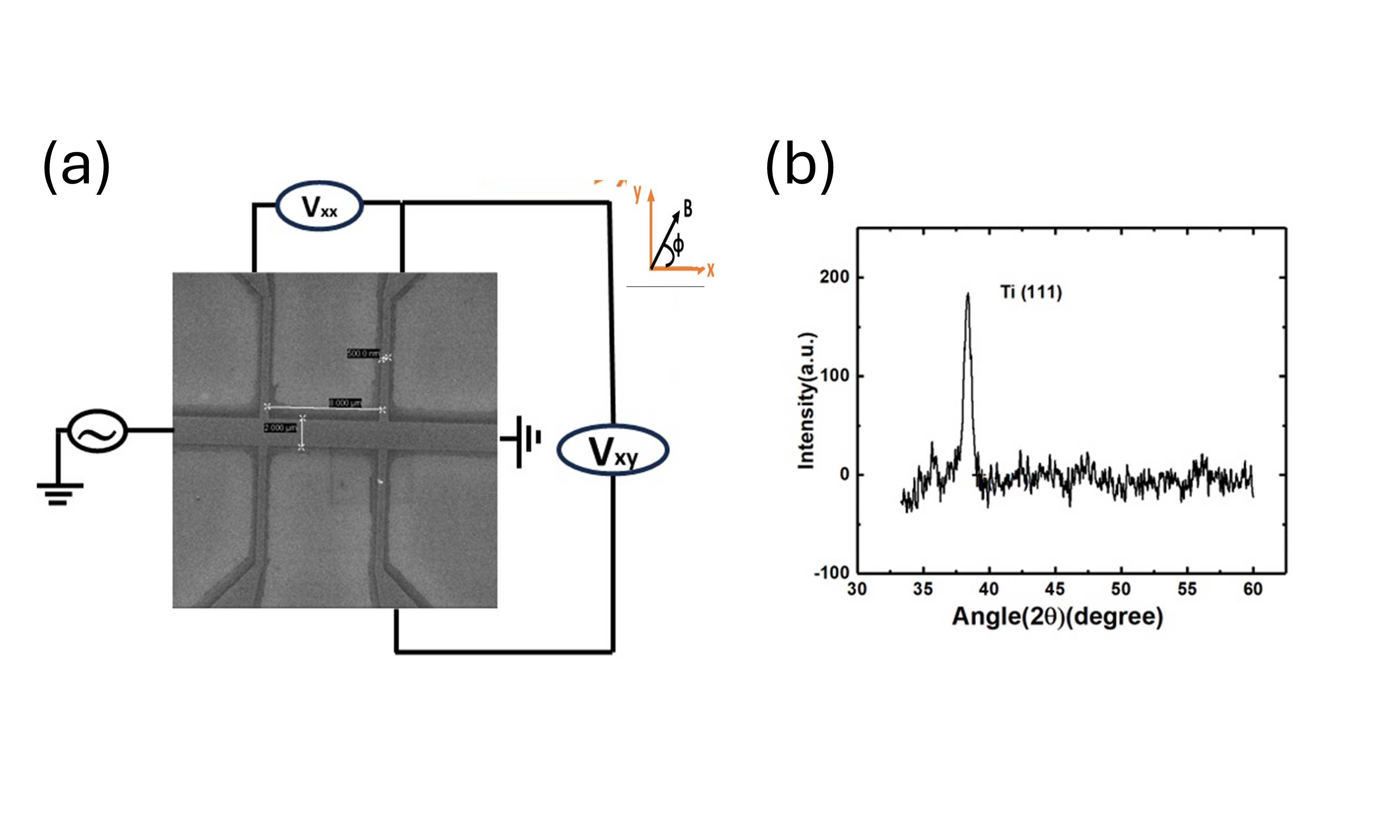}
 \caption{(a) Schematic of the Hall bar device geometry used for simultaneous measurements of longitudinal($V_{xx}$) and transverse ($V_{xy}$) voltages under an applied AC current. The inset shows an optical micrograph of the fabricated Hall bar. The external magnetic field (B) is applied at an in-plane angle ($\phi$) with respect to the current direction. (b) X-ray diffraction (XRD) pattern of the Ti layer showing a distinct (111) peak, indicating preferential crystalline orientation of the Ti film.}
 \label{schematic}
\end{figure}

\section {Torque of Cu/Ni and Cu/NiFe.}
\begin{figure}[h!]   
\includegraphics[width=0.5\textwidth]{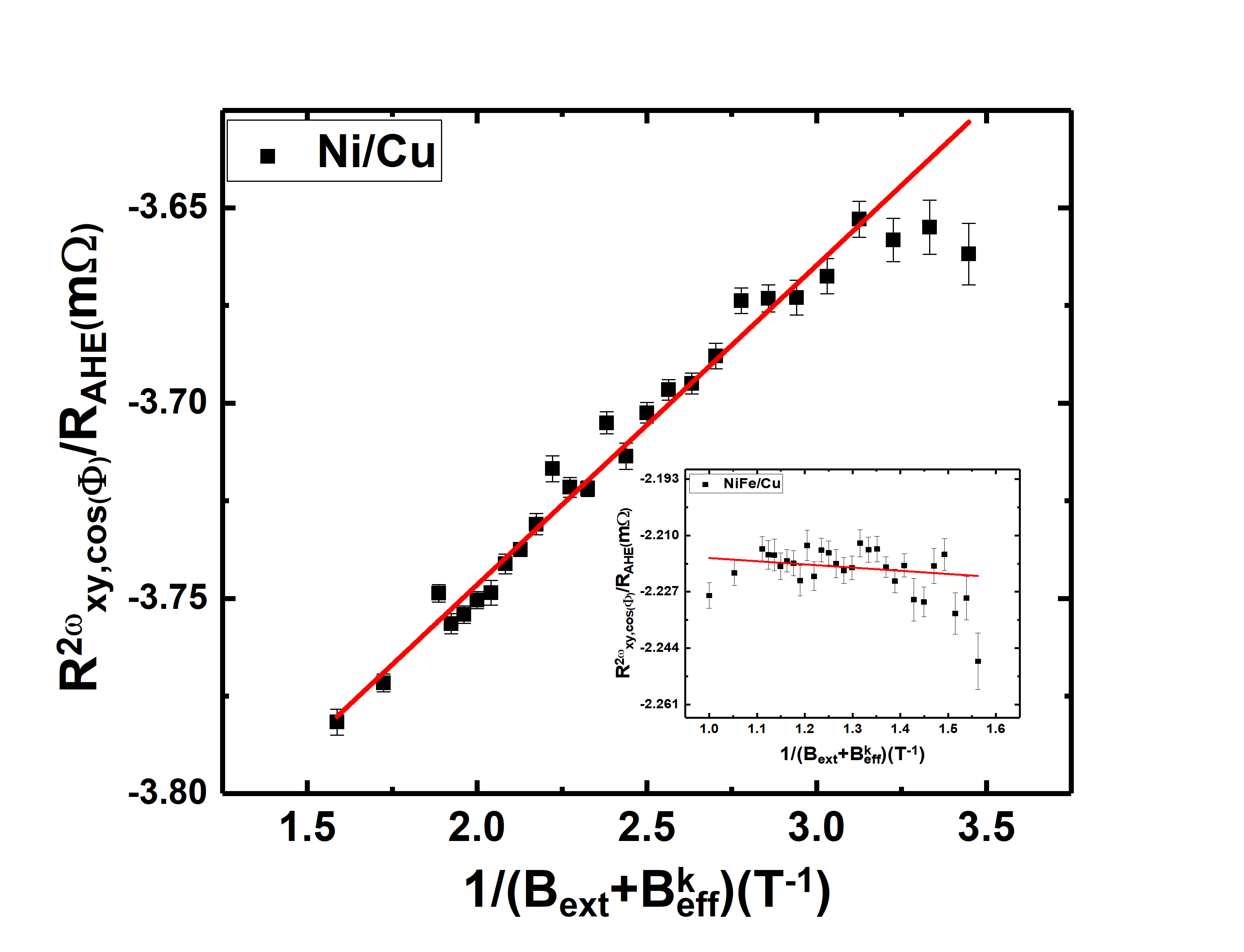}
 \caption{Normalized second-harmonic transverse resistance ($\frac{R_{xy, cos\phi}^{2\omega}}{R_{AHE}}$)  plotted as a function of $\frac{1}{B_{ext}+B^k_{eff}}$ for the Ni/Cu bilayer. The linear dependence (red line) indicates the presence of  SL-like torque contributions arising from the orbital Hall effect in Cu. The inset shows the corresponding measurement for NiFe/Cu, which exhibits a negligible slope, confirming the dominance of orbital-originated torque in the Ni/Cu system.}
 \label{schematic}
\end{figure}


\section{Electric field with temperature.}
\begin{figure}[h!]     
\includegraphics[width=0.4\textwidth]{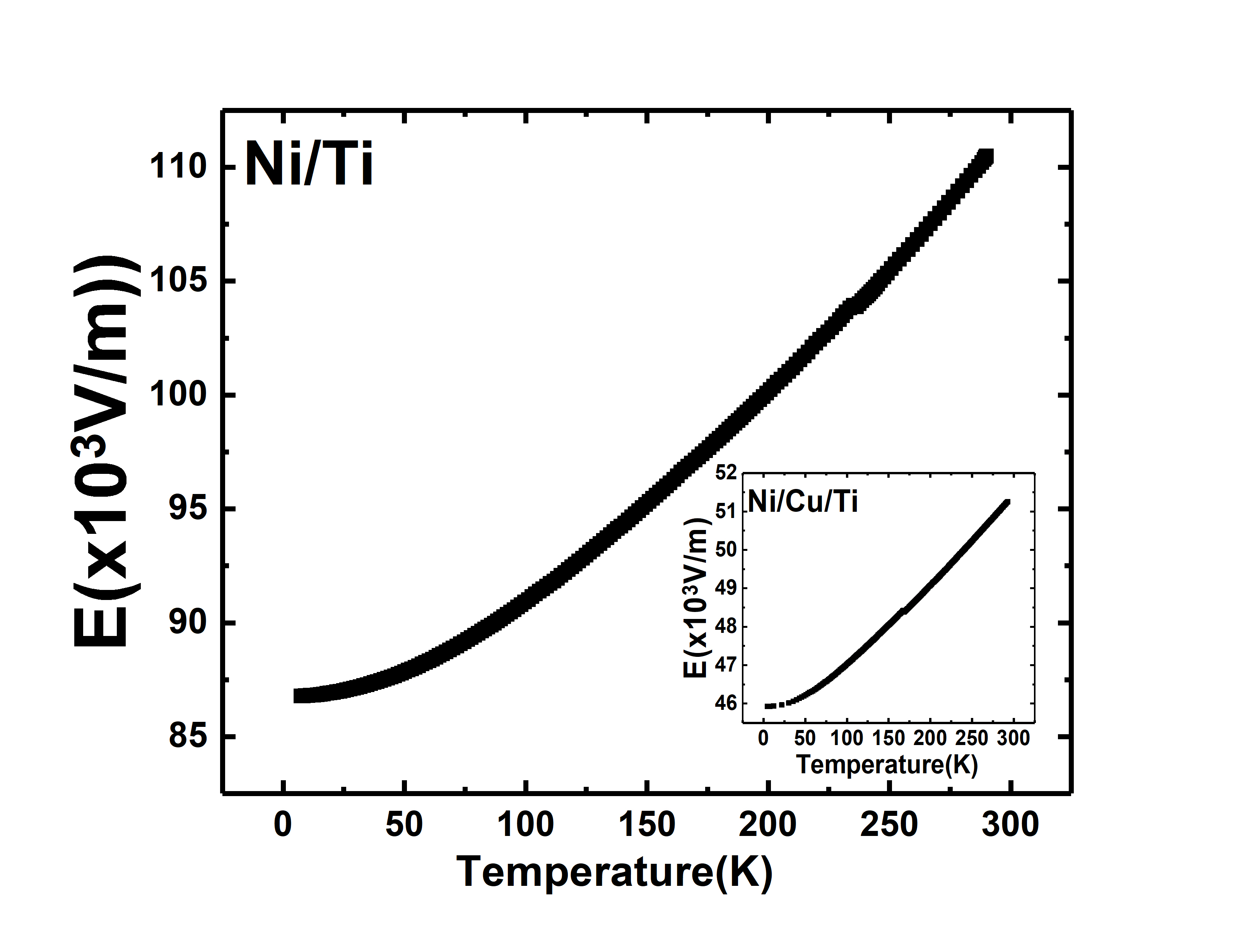}
 \caption{Temperature dependence of the applied electric field (E) for the Ni/Ti bilayer. The electric field increases gradually with temperature, indicating enhanced resistivity and reduced carrier mobility at higher temperatures. The inset shows a similar trend for the Ni/Cu/Ti trilayer, where a lower magnitude of E is observed due to the higher overall conductivity of the stack.}
 \label{schematic}
\end{figure}
\newpage
\section{AHE of all devices.}
\begin{figure}[h!]    
\includegraphics[width=0.5\textwidth]{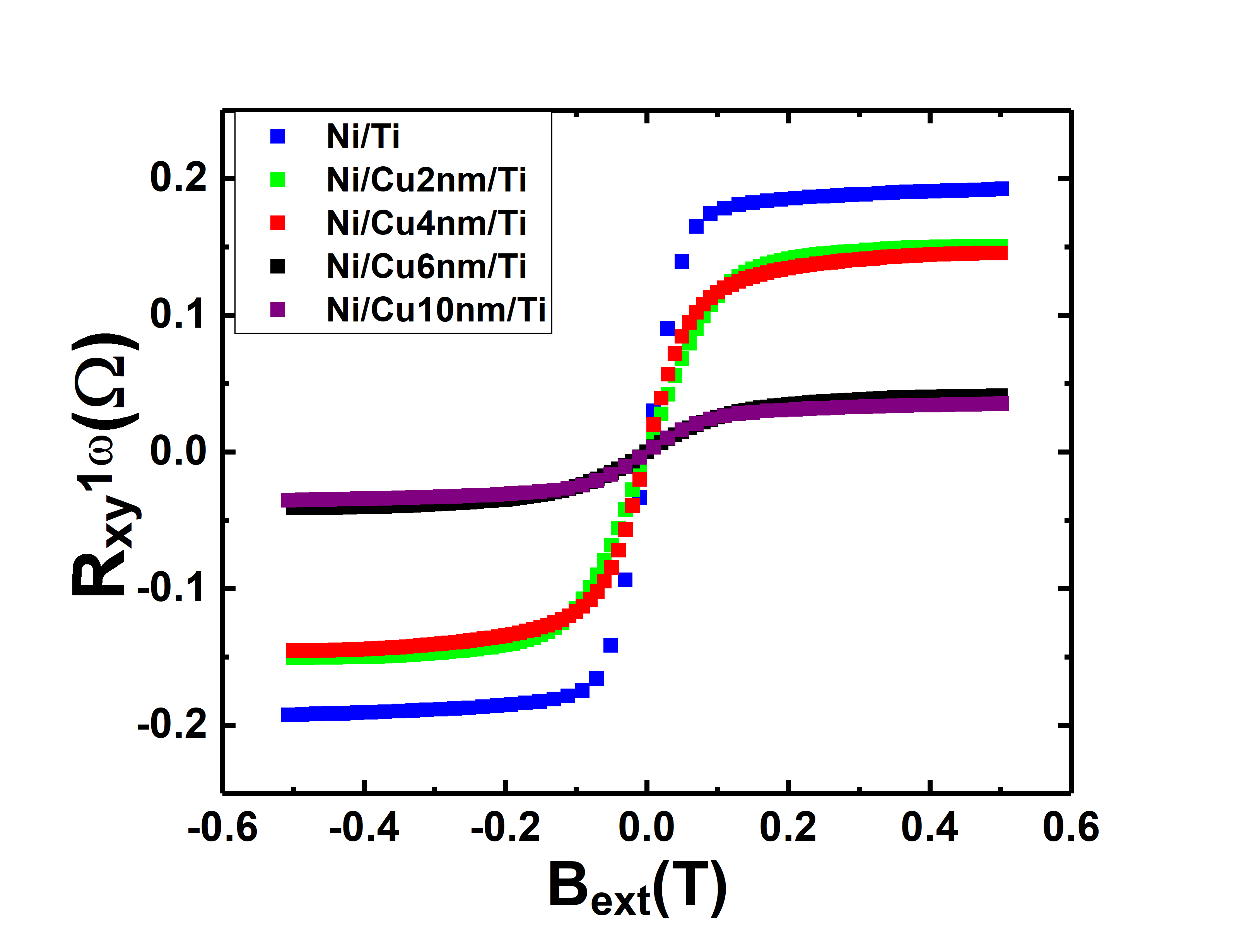}
 \caption{Out of plane magnetic field scan of Ni/Ti and Ni/Cu(t)/Ti devices.}
 \label{schematic}
\end{figure}

\nocite{*}
\newpage
\bibliography{apssamp}

\end{document}